\documentclass[preprint2]{aastex631}

\usepackage{hyperref}
\usepackage{textgreek}
\usepackage{graphicx}
\usepackage{physics}

\begin{document}

\title{Complete Sampling of the $uv$ Plane with Realistic Radio Arrays: Introducing the RULES Algorithm, with Application to 21 cm Foreground Wedge Removal}

\author[0000-0002-3317-5347]{Vincent MacKay}
\affiliation{MIT Kavli Institute, Massachusetts Institute of Technology, Cambridge, MA}

\author[0000-0001-5112-2567]{Zhilei Xu}
\affiliation{MIT Kavli Institute, Massachusetts Institute of Technology, Cambridge, MA}

\author[0000-0003-4980-2736]{Ruby Byrne}
\affiliation{Cahill Center for Astronomy and Astrophysics, California Institute of Technology, Pasadena, CA}

\author[0000-0002-4117-570X]{Jacqueline N. Hewitt}
\affiliation{MIT Kavli Institute, Massachusetts Institute of Technology, Cambridge, MA}
\affiliation{Department of Physics, Massachusetts Institute of Technology, Cambridge, MA}



\begin{abstract}

We introduce the Radio-array $uv$ Layout Engineering Strategy (RULES), an algorithm for designing radio arrays that achieve complete coverage of the $uv$ plane, defined as, at minimum, regular sampling at half the observing wavelength ($\lambda$) along the $u$ and $v$ axes within a specified range of baseline lengths. Using RULES, we generate $uv$-complete layouts that cover the range $10\lambda\leq\norm{\mathbf{u}}\leq 100\lambda$ with fewer than 1000 antennas of diameter $5\lambda$, comparable to current and planned arrays. We demonstrate the effectiveness of such arrays for mitigating contamination from bright astrophysical foregrounds in 21\,cm Epoch of Reionization observations---particularly in the region of Fourier space known as the \textit{foreground wedge}---by simulating visibilities of foreground-like sky models over the 130--150 MHz band and processing them through an image-based power spectrum estimator. We find that with complete $uv$ coverage, the wedge power is suppressed by sixteen orders of magnitude compared to an array with a compact hexagonal layout (used as a reference for a sparse $uv$ coverage). In contrast, we show that an array with the same number of antennas but in a random configuration only suppresses the wedge by three orders of magnitude, despite sampling more distinct $uv$ points over the same range. We address real-world challenges and find that our results are sensitive to small antenna position errors and missing baselines, while still performing equally or significantly better than random arrays in any case. We propose ways to mitigate those challenges such as a minimum redundancy requirement or tighter $uv$ packing density.

\end{abstract}

%\keywords{(...)}


\section{Introduction} \label{sec:intro}

In a radio interferometer, each antenna pair defines a baseline whose projected separation vector, in units of wavelength, corresponds to a sample in the so-called $uv$ plane. If the array is planar, the visibility function---sampled at each $uv$ coordinate---is Fourier conjugate to the sky intensity expressed in direction cosines, which reduce to angular coordinates in the flat-sky (small field) approximation \cite[pp.~767--781]{thompson2008interferometry}. Because the number of baselines is finite, the sampling function is necessarily incomplete: bounded, often sparse, and potentially non-uniform, leading to artifacts in the reconstructed sky image, including point spread function (PSF) sidelobes, aliasing, and edge effects. The resulting map is called a \textit{dirty image}, and the associated PSF, the \textit{dirty beam}.

Several strategies have been developed to mitigate the effects of incomplete $uv$ sampling.
Some techniques increase coverage by leveraging the temporal and spectral axes: Earth rotation synthesis takes advantage of the changing projection of baselines as the Earth rotates over time, while multifrequency synthesis uses the frequency dependence of baseline lengths—combined with the assumption that sources have a smooth spectral structure \citep[pp.~31--34, 578--579]{thompson2008interferometry}. Deconvolution approaches seek to correct for incomplete sampling by iteratively modeling and subtracting sources to suppress PSF-induced distortions in the image. These include CLEAN \citep{CLEAN:1974Hogbom,schwab1984,Cotton2004,cotton2005}, A-projection \citep{aproj2008,carozzi2009}, forward modeling \citep{forward_modeling}, and fast holographic deconvolution (FHD, \citealt{FHD2012}); for a comprehensive overview, we refer the reader to \cite{FHD2012}. Some arrays address the problem structurally through reconfigurable layouts: antennas can be moved between fixed “pads” to realize distinct and complementary configurations. These include linear rails (e.g., the Synthesis Telescope, \citealt{synthesis_tel_2000}), T- or Y-shaped tracks (e.g., the VLA, \citealt{VLA1980}), or custom transporters that enable arbitrary moves (e.g., ALMA, \citealt{alma2004}). Alternative array designs aim to maximize distinct $uv$ samples from the outset: configurations with offset sub-arrays have been proposed for this purpose \citep{Dillon_Parsons_Redundant_Configs_2016}, as have layouts inspired by Golomb rulers---mathematical constructs in which all pairwise differences between elements are distinct (\citealt{bireau1974golomb}; \citealt[pp.~173--174]{thompson2008interferometry}; \citealt{Parsons_golomb}; \citealt{ebrahimi_may2023}; \citealt{lazko_nov2023})---sometimes obtained via algorithms for optimal antenna placement \citep{Keto_1997,Boone2001,Boone2002,cohanim2004,MurrayTrott2018Dec}. Meanwhile, large arrays increasingly adopt random or pseudo-random layouts to achieve relatively uniform $uv$ coverage (e.g., MeerKAT, \citealt{meerkat2009}; MWA, \citealt{MWA_design_overview}; DSA-2000, \citealt{DSAHallinan2019Sep}; SKA, \citealt{SKAWeltman2020Jan}).

One specific challenge that follows from incomplete $uv$ sampling, and that is particularly severe in 21\,cm cosmology, is spectral leakage of bright foregrounds in power spectrum estimates. This contamination is most important in the so-called \textit{foreground wedge}, a region of the two-dimensional power spectrum at low $k_\parallel$, where the instrument’s intrinsic chromaticity and imperfect calibration causes foregrounds to spill into the cosmological signal window (see \autoref{sec:wedge}, and \citealt{Bowman2009Mar,Datta2010Nov,Morales2012Jun,Parsons2012Aug}). The wedge not only reduces sensitivity to the 21\,cm signal, but hinders cross-correlation with other surveys, prevents most imaging-based analyses \citep{Pober2014ApJ,beardsley_2015,cross_cor_wedge_2016,cross_corr_cohn_2016,cross_cox_2022,gagnon_hartman_2024}, notably one-point statistics \citep{kittiwisit2017,Kim2025_onepoint}, and affects calibration negatively \citep{calib_barry_2016,calib_ewallwice_2017,calib_Byrne_2019}. While many existing methods to tackle imperfect $uv$ sampling, such as those enumerated above, are effective at improving source localization or imaging fidelity, only some are equipped to address the foreground wedge. Pseudo-random layouts, for instance, are sometimes adopted to reduce leakage by lowering redundancy and spreading out $uv$ coverage, but their performance is not guaranteed; they may still leave gaps or other artifacts caused by uneven sampling, and their effectiveness is difficult to predict and control. Techniques based on wedge subtraction also exist \citep{liutegmark2012,paciga2013,Liu2014b,mertens2018,Cox2024Nucal}, but have yet reached levels required by 21\,cm science, and are often constrained by the limitations of the underlying $uv$ sampling (see \citealt{liushaw2020} for a review).

Given the fact that foregrounds and calibration systematics dominate the error budget on most baselines in current and planned 21\,cm arrays---they are not yet noise limited---and that nearly all of the strategies for improved $uv$-sampling listed previously were developed for arrays with $\lesssim$\,100 antennas, it is relevant to explore new design-based approaches suited to the emerging $\mathcal{O}(10^3)$-antenna era \citep{Vanderlinde2019Oct,DSAHallinan2019Sep,SKAWeltman2020Jan}. The scale of these modern instruments opens the door to arrays that sample the $uv$ plane more systematically, offering the potential to fully suppress the wedge through layout geometry alone, an idea that has been presented in \cite{MurrayTrott2018Dec}.

In this work, we explore whether it is possible to outperform random layouts by constructing antenna configurations that deliberately realize a dense and regular $uv$ sampling function. This parallels the approach of \cite{MurrayTrott2018Dec}, where the authors propose a logarithmic $uv$ distribution; in contrast, we argue for a square $uv$ lattice, based on sampling theory, and the fact that the finite extent of the sky translates to a maximum spatial frequency in the $uv$ plane. We further show that such arrays are feasible under realistic parameters, and introduce the Radio-array $uv$ Layout Engineering Strategy (RULES), an algorithm that generates these layouts based on user-chosen constraints.

The rest of this paper is organized as follows. In \autoref{sec:completeness_criterion}, we formalize the completeness criterion. In \autoref{sec:designing_the_array}, we present the RULES algorithm. In \autoref{sec:wedge}, we evaluate the performance of an algorithmically-generated, ``$uv$-complete'' array in terms of 21\,cm foreground suppression, compared to a regular and a random array, and address the question of feasibility. Finally, in \autoref{sec:discussion}, we discuss the benefits of $uv$ completeness further and the potential application to high-resolution imaging beyond 21\,cm science, and propose future work, and we conclude in \autoref{sec:conclusion}.

%The following papers have argued that a dense $uv$ coverage could mitigate the wedge: \cite{Bowman2009Mar}, \cite{Morales2012Jun}, and \cite{Parsons2012Aug}.

\section{Completeness criterion}
\label{sec:completeness_criterion}

From the van Cittert–Zernike theorem, the sky intensity---projected to direction cosine coordinates---and the $uv$ plane from a flat radio array constitute a Fourier conjugate pair \cite[pp.~767--781]{thompson2008interferometry}. The visibility function in the $uv$ plane is spatially band-limited, with a maximum spatial frequency of $1/\lambda$ (where $\lambda$ is the observing wavelength), which occurs when the source lies in a direction such that the geometric delay is maximal---that is, when the source direction is parallel to the baseline, as with horizon sources. According to sampling theory, this band-limited nature ensures that perfect reconstruction is possible from discrete measurements, provided the sample spacing is regular, small enough---specifically, no greater than $\lambda/2$---and infinite in extent (\citealt{gasquet1998fourier}, pp.~355--357; \citealt{gray2012fourier}, pp.~327--331). While physical arrays necessarily have finite extent, they can still achieve the required sampling regularity and density over a bounded region of the $uv$ plane, which is sufficient to substantially reduce imaging artifacts and spectral leakage. To quantify the sampling density, we define a parameter $\rho$ such that, for a square lattice of $uv$ points, $\rho$ is the inverse of the lattice spacing in units of $\lambda$; in other words the sample spacing along either axis of the $uv$ plane is $\lambda/\rho$. An array that realizes a regular $uv$ grid thus meets the sampling density criterion---and is therefore said to be \textit{$uv$ complete} within a given baseline range $u_\mathrm{min}\leq\norm{\mathbf{u}}\leq u_\mathrm{max}$---if it has $\rho \geq 2$.

Since 21\,cm observatories typically span a wide frequency band, using a single reference wavelength $\lambda$ to define $uv$ density is imperfect. In this work, we adopt the \textit{shortest} wavelength in the band as our reference, ensuring that the $\lambda/2$ sampling criterion is satisfied across the full bandwidth, but this choice implies that the completeness bounds $u_\mathrm{min}$ and $u_\mathrm{max}$, expressed in units of $\lambda$, will shift with frequency.

\begin{figure}[ht]
    \centering
    \includegraphics[width=\linewidth]{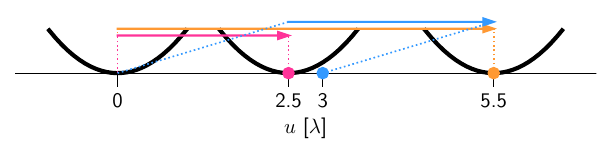}
    \caption{Illustration of the one-dimensional toy example showing how $uv$ points can be closer than the antenna size. Antennas of size $2\lambda$ located at $0\lambda$, $2.5\lambda$, and $5.5\lambda$ form $uv$ points at $2.5\lambda$, $3\lambda$, and $5.5\lambda$. This paper uses this approach, generalized to the 2D plane, to obtain a $\lambda/2$ $uv$ coverage with antennas of size $D>\lambda$.}
    \label{fig:toy_example}
\end{figure}

This target $uv$ density is realizable even if $\lambda/2$ is smaller than the antenna diameter. We illustrate how this is possible with a simple one-dimensional example: consider three antennas of size $D = 2\lambda$ placed at $0\lambda$, $2.5\lambda$, and $5.5\lambda$. The resulting $u$ coordinates are $2.5\lambda$, $3\lambda$, and $5.5\lambda$, two of which are separated by $\lambda/2$. This toy layout, shown in \autoref{fig:toy_example}, generalizes to two dimensions and allows dense sampling even with $D > \lambda/2$. There will necessarily remain a gap where $\norm{\mathbf{u}} < D$, and many of the formed baselines may sample identical $uv$ points or lie outside the range of $uv$ completeness.

Throughout this work, we restrict our analysis to planar arrays and set $w = 0$. Since the sky is two-dimensional, its Fourier conjugate space can be fully sampled by a planar array, making nonzero $w$ components unnecessary in principle. In practice, mechanical or site constraints may introduce small $w$ values, but fully accounting for them requires more complex imaging techniques---such as $w$-projection or $w$-stacking---which, while increasingly tractable thanks to advances in high-performance computing \citep{w_proj_lao_2019,w_stack_gheller_2023}, introduce additional complications that we choose to leave out of the scope of this paper. Likewise, while space-based arrays may offer sampling advantages through unconstrained three-dimensional baselines, we do not consider them here, as current space-based proposals remain far from achieving the $\mathcal{O}(10^3)$-element, tightly packed layouts feasible on Earth \citep{space_based_array_2010,space_based_array_2016,space_based_array_2020}. We note, however, that such arrays still measure a visibility function that is limited to a maximum spatial frequency of $1/\lambda$ in the $uvw$ space.

%A complementary perspective comes from diffraction theory. The relation between sample spacing $d$ and the angular position $\theta$ of the $m$\textsuperscript{th} grating null is given by $d \sin(\theta) = \left(m + \tfrac{1}{2}\right)\lambda$. Pushing the first grating null ($m = 0$) to the horizon ($\sin(\theta) = 1$) requires $d = \lambda/2$, consistent with the sampling criterion above.

%We note that our completeness criterion is conceptually related to Golomb rulers—sets of non-negative integers with all pairwise separations distinct—and to $m$-perfect rulers, which guarantee that every integer $k \leq m$ is uniquely expressible as a difference between two elements. While these constructs have been studied in the context of radio astronomy, particularly for one-dimensional arrays, our problem differs in several key respects:
%\begin{enumerate}
%    \item We operate in two dimensions, whereas Golomb and perfect rulers are typically defined on a line.
%    \item Golomb rulers prioritize uniqueness of separations, not completeness; perfect rulers aim for both. Our criterion instead requires completeness (i.e., all desired separations are realized) but imposes no constraint on uniqueness.
%    \item Antenna placements must obey a minimum separation constraint, typically larger than $\lambda/2$, unlike ruler constructions where adjacent marks are allowed.
%\end{enumerate}

\section{Designing the array with RULES}\label{sec:designing_the_array}

Our method starts from a set of $uv$ points that we wish to sample---the \textit{commanded baselines}---which are then \textit{fulfilled} by iteratively adding antennas at carefully picked positions. The chosen set of commanded baselines is justified in \autoref{subsec:commanded_baselines}, and the antenna-generating algorithm---RULES---is presented in \autoref{subsec:the_algorithm}.

\subsection{The commanded baselines}\label{subsec:commanded_baselines}

\begin{figure}[ht]
    \centering
    % uncomment the next line, and comment the following one, if the figure encoraches on the text
    %\adjustbox{padding=0ex 0ex 0ex 2ex}{\includegraphics[width=\linewidth]{commanded_lamdba.pdf}}
    \includegraphics[width=\linewidth]{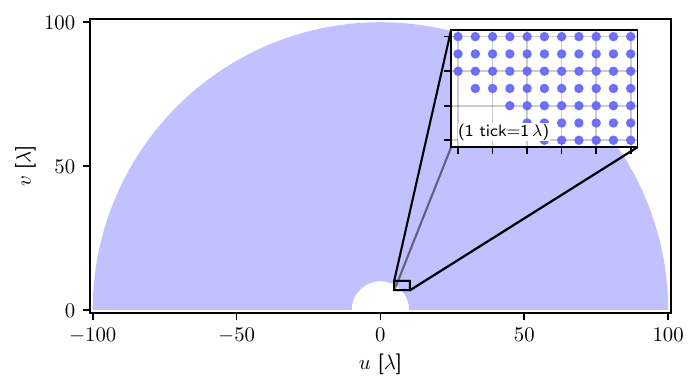}
    \caption{Commanded baselines $U_\mathrm{C}$, arranged in a square grid with a density of $\lambda/2$ over $10\lambda\leq \norm{\mathbf{u}}\leq 100\lambda$; only half of the full annulus is shown since $\mathbf{u}$ and $-\mathbf{u}$ are the same baseline.}
    \label{fig:commanded_lambda}
\end{figure}

The commanded baselines form a square lattice with $\lambda/2$ spacing. This regular grid is motivated by sampling theory (see \autoref{sec:completeness_criterion}) and has the added benefit of generating a discretized aperture plane, which enables fulfillment of all baselines with relatively few antennas, and leads to exact baseline redundancy (see \autoref{subsec:redundancy}). For this work, we adopt a baseline length range of $u_\mathrm{min} = 10\lambda$ to $u_\mathrm{max} = 100\lambda$, though RULES supports other choices (see \autoref{apsec:parsweeps}). We reiterate that $\lambda$ is defined as the shortest wavelength in the observing band; this ensures the desired $uv$ density across all frequencies, but means that the completeness range, expressed in wavelengths, will fall short of 100 at longer wavelengths. The number of commanded baselines is $N_\mathrm{C} = 62{,}186$, and the set is defined as $U_\mathrm{C} = \{\mathbf{u_1}, \mathbf{u_2}, \ldots, \mathbf{u}_{N_\mathrm{C}}\}$, shown in \autoref{fig:commanded_lambda}.

\subsection{The RULES algorithm}
\label{subsec:the_algorithm}

Let $A$ be the set of antennas in the array, starting with a single antenna, i.e.,  $A=\{(0,0)\}$. At each iteration, a reference antenna $\mathbf{a}_\mathrm{ref} \in A$ and a commanded baseline $\mathbf{u}_\mathrm{C} \in U_\mathrm{C}$ are selected to generate a candidate position $\mathbf{a}_\mathrm{new} = \mathbf{a}_\mathrm{ref} + \mathbf{u}_\mathrm{C}$. A collision occurs if this candidate position is closer than some distance $D$ to any other antenna in $A$. While $D$ is typically chosen to be the antenna's physical size, it can be set to a larger value to enforce minimum spacing for other considerations such as reducing mutual coupling. If there is no collision, $\mathbf{a}_\mathrm{new}$ is added to $A$, and $\mathbf{u}_\mathrm{C}$ is removed from $U_\mathrm{C}$. If there is a collision, the mirrored position $\mathbf{a}_\mathrm{ref} - \mathbf{u}_\mathrm{C}$ is tried. If both fail, a new $\{\mathbf{a}_\mathrm{ref},\mathbf{u}_\mathrm{C}\}$ pair is chosen. The process continues until all commanded baselines are fulfilled or additional constraints---such as a maximum number of antennas or array size—halt the algorithm.

\begin{figure}[ht]
    \centering
    \includegraphics[width=\linewidth]{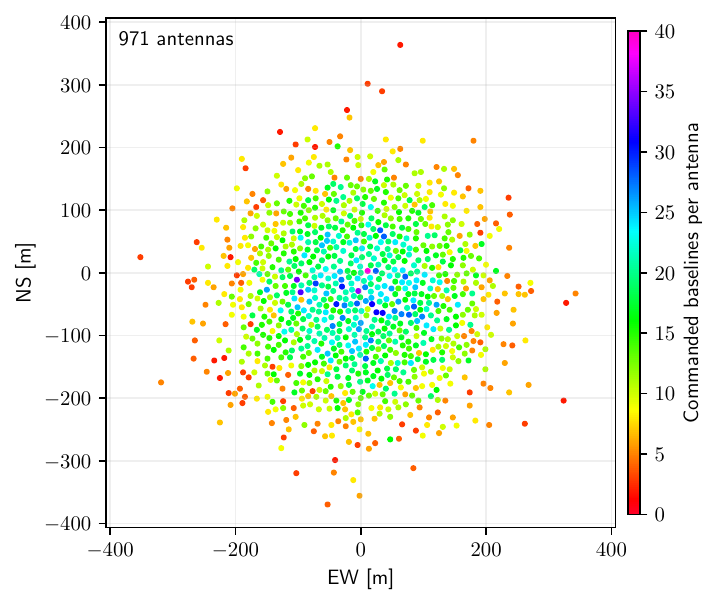}
    \caption{Array produced with the RULES algorithm presented in \autoref{subsec:the_algorithm}, fulfilling all the commanded baselines from \autoref{subsec:commanded_baselines} shown in \autoref{fig:commanded_lambda}. To generate it, the next commanded baseline to fulfill was chosen according to a predetermined order---shortest to longest---while the next reference antenna was chosen by comparing all existing antennas and picking the one that would fulfill the most commanded baselines. Arrays generated with other parameters are shown in \autoref{apsec:parsweeps}.}
    \label{fig:array_layouts}
\end{figure}

At first glance, one might expect RULES to generate an array of $N_\mathrm{C}$ antennas---one per commanded baseline. In practice, however, since the commanded baselines lay on a grid, each new antenna typically fulfills several points in $U_\mathrm{C}$ beyond the chosen $\mathbf{u}_\mathrm{C}$. These coincidentally fulfilled baselines are also removed from $U_\mathrm{C}$, allowing the algorithm to complete with far fewer than $N_\mathrm{C}$ antennas. The number of antennas depends critically on how each $\{\mathbf{a}_\mathrm{ref}, \mathbf{u}_\mathrm{C}\}$ pair is selected. Using a fixed order for both yields fast runtimes (10--20 minutes on a single-threaded laptop) but typically requires 1800--3000 antennas to fulfill all baselines---depending on the order. At the other extreme, evaluating all possible combinations of $\mathbf{a}_\mathrm{ref}$ and $\mathbf{u}_\mathrm{C}$ at each step, and choosing the one that fulfills the most commanded baselines, produces the most compact layout (938 antennas, shown in \autoref{fig:extra_arrays}a), but comes at steep computational cost: nearly 40 hours on a 64-core cluster. We find an effective compromise by fixing the order of the $\mathbf{u}_\mathrm{C}$'s (shortest to longest seems to work best) while only comparing the $\mathbf{a}_\mathrm{ref}$ candidates at each step. This hybrid strategy produced a layout with only 971 antennas (\autoref{fig:array_layouts}) in under 20 minutes using 12-core parallelization on a modern laptop, and allows additional optimization criteria such as compactness or spacing to be incorporated with minimal overhead. Apart from \autoref{fig:extra_arrays}a, all RULES-based arrays mentioned in this paper use the hybrid approach.

An important feature of RULES is that when comparing multiple $\{\mathbf{a}_\mathrm{ref},\mathbf{u}_\mathrm{C}\}$ pairs, if one results in a collision or is prevented by some other constraint, it is noted and skipped at the next iteration. This pruning significantly decreases completion time.

We emphasize that the figures quoted above apply specifically to the set of commanded baselines described in \autoref{subsec:commanded_baselines}, with a collision constraint of $5\lambda$ and no maximum array size; in general, the computation time and number of antennas required to complete the array are a function of the set of commanded baselines and imposed constraints. For example, one may want to increase the distance between antennas to reduce mutual coupling---this is possible, as the $uv$ completeness is achieved through density in the \textit{uv} plane, not in the aperture plane---but will require more antennas. In \autoref{apsec:parsweeps}, we present how RULES performs with other parameters. We also stress that the lattice nature of the commanded points is favorable to RULES, whereas random, pseudo-random, or other sets of commanded points that do not lay on a grid require many more antennas because they do not produce the same rate of coincidental fulfillments.

These results demonstrate that it is possible to fulfill a regular and densely sampled $uv$ coverage using a number of antennas comparable to instruments that are currently under development such as CHORD \citep{Vanderlinde2019Oct}, DSA-2000 \citep{DSAHallinan2019Sep}, and the SKA \citep{SKAWeltman2020Jan}; $uv$ completeness is thus achievable within practical design constraints. We have implemented the RULES algorithm in a Python package, accessible publicly on GitHub.\footnote{\url{https://github.com/vincentmackay/uvrules}}

\subsection{Discretized aperture plane and redundancy}\label{subsec:redundancy}

\begin{figure}
    \centering
    \includegraphics[width=\linewidth]{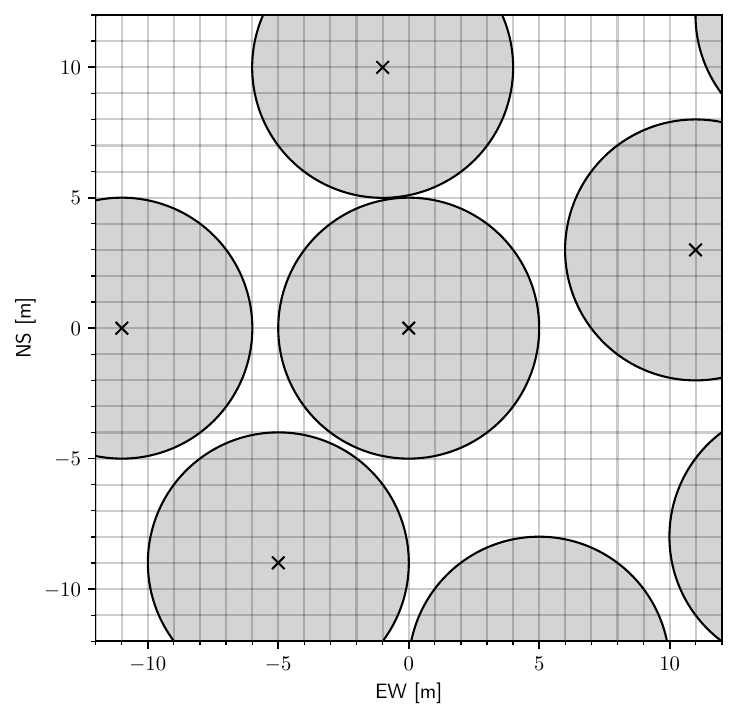}
    \caption{Aperture plane for the RULES-based array  presented in \autoref{fig:array_layouts}, zoomed around $(0,0)$ to show that the centers (crosses) of antennas (gray disks) fall on a fine grid of size $\lambda/2 = 1\,\mathrm{m}$, leading to exact redundancy (down to the antenna position error $\sigma_\mathrm{pos}$).}
    \label{fig:a971_zoomed}
\end{figure}

A fundamental tension exists between $uv$ completeness and redundancy. Redundancy has benefits such as noise reduction through coherent averaging, increased tolerance to missing antennas, and the possibility of decreasing data volume by combining identical baselines. However, for a fixed number of antennas, attempting to cover more of the $uv$ plane inherently limits the number of redundant $uv$ points, and vice versa. RULES provides a compromise: by design, all antennas end up at integer multiples of $\lambda/2$ along the EW and NS directions, such that the aperture plane is effectively discretized---a feature that is illustrated in \autoref{fig:a971_zoomed}. Consequently, multiple baselines are bound to coincide \textit{exactly}---down to the antenna position error $\sigma_\mathrm{pos}$---leading to a higher level of redundancy than a random array of a similar size.

\begin{figure}
    \centering
    \includegraphics[width=\linewidth]{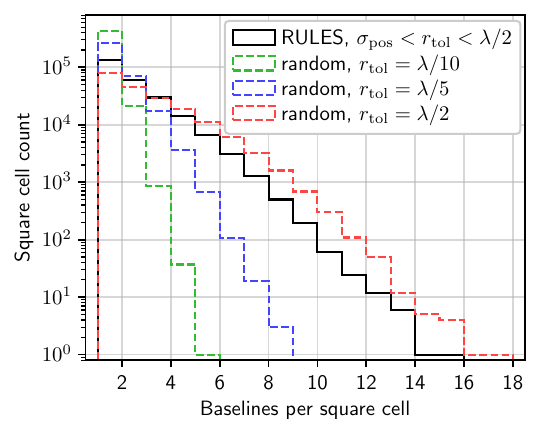}
    \caption{Redundancy count, defined by dividing the $uv$ plane in $r_\mathrm{tol}$-sized square cells and counting the number of baselines per square cell; baselines occupying the same cell are considered redundant. An array generated with the algorithm (\texttt{RULES}, as in \autoref{fig:array_layouts}) results in higher redundancy than a comparable randomly generated array (\texttt{random}, which has the same number of antennas and physical footprint as \texttt{RULES}), until $r_\mathrm{tol}$ is loosened to $\lambda/2$. The redundancy values for $\texttt{RULES}$ are identical for all values of $r_\mathrm{tol}<\lambda/2$ (down to the antenna position error $\sigma_\mathrm{pos}$) because its $uv$ samples lie on a $\lambda/2$ square grid.}
    \label{fig:redundancy}
\end{figure}

To demonstrate this feature, we compare the array presented in \autoref{fig:array_layouts} (labeled \texttt{RULES}) with a comparable random array (\texttt{random}) shown in \autoref{fig:array_uv_beam} (top center---names in monospaced typeface henceforth refer to the arrays presented in that figure), which has the same number of antennas and a similar physical footprint. We define the redundancy metric by dividing the $uv$ plane in a lattice of square cells of side length $r_\mathrm{tol}$ (the \textit{redundancy tolerance}); baselines occupying the same cell are considered redundant. In \autoref{fig:redundancy}, the redundancies are shown for both arrays for different values of $r_\mathrm{tol}$. For \texttt{RULES}, since the $uv$ points lie on a $\lambda/2$ grid, the redundancy remains the same at all values of $r_\mathrm{tol}<\lambda/2$, with $\mathcal{O}(10^3)$ cells containing 5 to 10 redundant baselines. In comparison, \texttt{random} needs $r_\mathrm{tol}\sim \lambda/2$ to reach similar numbers, and most cells have only one baseline at $r_\mathrm{tol}\leq 0.1\lambda$.

We consider the increased redundancy a fortuitous property of RULES, and leave the full analysis of those benefits out of the scope of this paper. We also note that the discretized nature of the aperture plane may help with precisely positioning the antennas when building the array. We however acknowledge that a RULES-based array remains much less redundant than a standard regular-lattice array with the same number of antennas, such as the hexagonal realization presented in \autoref{fig:array_uv_beam} (top left), which can have redundancies reaching $\mathcal{O}(N_\mathrm{A})$---where $N_\mathrm{A}$ is the number of antennas---at vanishing values of $r_\mathrm{tol}$.


\section{The 21\,cm foreground wedge}\label{sec:wedge}

\begin{figure*}[t]
    \centering
    \includegraphics[width=\textwidth]{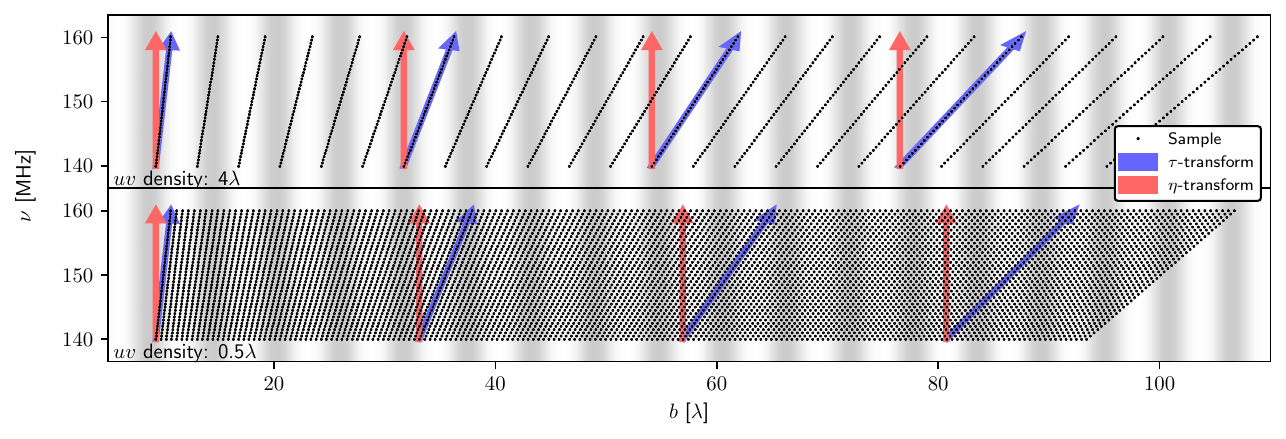}
    \caption{Shaded background: spatial modes (real part, one dimension) of a flat-spectrum point source, $10^\circ$ from field center. Black dots are sampled $uv$ points, at a $4\lambda$ and $\lambda/2$ spacing in the top and bottom panels, respectively. Blue arrows are delay ($\tau$-)transforms, and red arrows are line-of-sight wavenumber ($\eta$-)transforms.}
    \label{fig:undersampling}
\end{figure*}

The foreground wedge is a common feature in 2D power spectra of radio interferometric measurements. It represents power from the spectrally smooth and very bright foregrounds leaking at low $k$'s---and higher $k_\parallel$ with increasing $k_\perp$---partially masking the much fainter neutral hydrogen signal. The emergence of the wedge and its connection to incomplete $uv$ sampling can be derived in various ways, and we direct the reader to \cite{Bowman2009Mar}, \cite{Datta2010Nov}, \cite{Morales2012Jun}, and \cite{Parsons2012Aug} for deeper reviews. We summarize here a simple heuristic, using \autoref{fig:undersampling}, which shows the visibility (real part, collapsed to one dimension) of a flat-spectrum point source located $10^\circ$ from field center, with black dots representing the sampled $uv$ points.

To produce a power spectrum, this visibility must be Fourier transformed along the line-of-sight axis; since the foregrounds are smooth in that direction, the power should be concentrated at low modes. In practice, that is usually not what happens, and power leaks to higher modes at longer baselines, constituting the wedge. The blue arrows in \autoref{fig:undersampling} correspond to per-baseline Fourier transforms---known as \textit{delay} transforms, or $\tau$-transforms \citep{Parsons2012Aug}---which produce the wedge at any $uv$ density as the transform axes are not parallel to the line-of-sight axis, and cross more and more spatial mode crests with increasing baseline lengths. We note that wedgeless $\tau$-transforms might be achievable if foregrounds are preliminarily subtracted, but no analysis method has yet reached the subtraction precision required for 21\,cm cosmology. The red arrows represent Fourier transforms in an image-based power spectrum pipeline \citep{Trott2016Feb,Patil2017Mar,Barry2019Jan,Xu2024Aug}, sometimes referred to as the $\eta$-transform (where $\eta$ is the line-of-sight wavenumber). That framework amounts to aligning the sampled visibilities in $uv$ space, or the pixels in image space, such that the transform along the frequency axis does not drift over multiple spatial modes. If the $uv$ samples are too sparse, this cannot work for long baselines because the samples are too far from the transform axis, such that the aligned samples must be interpolated, with artifacts also leading to a wedge. However, if the $uv$ plane is densely sampled, this issue can be mitigated. This is demonstrated in \autoref{subsec:ps_results}, using a RULES-based array and the Direct Optimal Mapping power spectrum (DOM-PS) pipeline \citep{Xu2022Oct,Xu2024Aug}; the result is repeated with the FHD/$\varepsilon$ppsilon estimator \citep{Barry2019Jan}, in \autoref{apsec:eppsilon}.

\begin{figure}
    \centering
    \includegraphics[width=\linewidth]{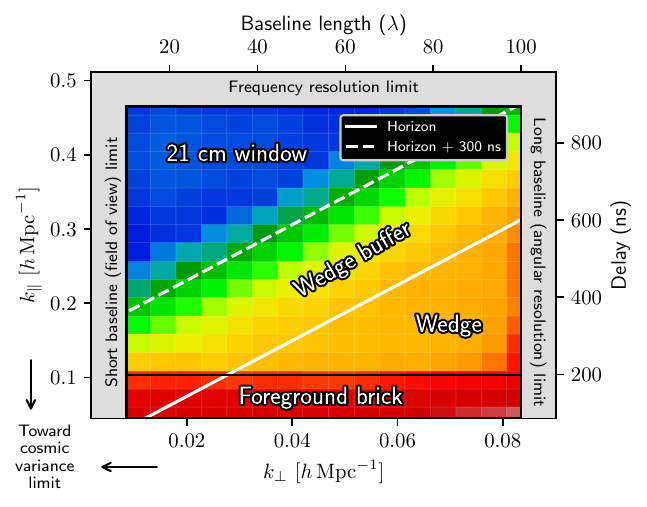}
    \caption{Schematic diagram of the foreground wedge, superimposed on a simulated two-dimensional power spectrum (from \autoref{fig:pspecs}, top left). The foreground brick at the bottom represents chromatic leakage from bright foregrounds due to the finite bandwidth. The wedge is below the horizon delay line, with a wedge buffer also accounting for the bandwidth. The 21\,cm window, free of foreground contamination, is in the top-left. Gray regions indicate instrumental limits in $k_\perp$ and $k_\parallel$, while arrows mark the direction toward the cosmic variance limit at low $k$.}
    \label{fig:wedge_cartoon}
\end{figure}

\autoref{fig:wedge_cartoon} shows a schematic of the foreground wedge, superimposed on a simulated 2D power spectrum (from \autoref{fig:pspecs}, top left). At the bottom is the foreground brick, a band of low-$k_\parallel$ power from intrinsic foreground chromaticity that leaks to higher $k_\parallel$ due to the finite-domain Fourier transform along the line-of-sight axis. While this leakage typically scales roughly as the inverse bandwidth ($\sim$\,50\,ns), it is bigger here due to the window function we used for the Fourier transform (7-term Blackman-Harris), which provides a better dynamic range but a wider main lobe. The wedge appears below the horizon delay line, with a buffer added above it, also caused by the finite bandwidth. The 21\,cm window occupies the top-left region. Instrumental limits truncate the power spectrum at low $k_\bot$ (field of view, set by the shortest baseline), high $k_\bot$ (angular resolution, set by the longest baseline), and high $k_\parallel$ (frequency resolution, set by the backend hardware). There is no instrumental limit at low $k_\parallel$, though the bandwidth determines the size of the brick, and cosmic variance eventually dominates.

\subsection{Simulation and analysis pipeline}\label{subsec:sim_setup}

Visibilities used in this section are produced with the \texttt{pyuvsim} simulator \citep{pyuvsim_Lanman2019}, using GLEAM \citep{GLEAM} for point sources and GSM08 \citep{GSM08} for the diffuse sky, which we show separately. The frequency range is 130--150\,MHz---chosen as mid-band for Epoch of Reionization (EoR) experiments---and the reference wavelength used for RULES is the shortest wavelength of the band, $\lambda = 2\,\mathrm{m}$. The beam model is the Airy pattern associated with a $D = 5\lambda = 10\,\mathrm{m}$ aperture. Simulations include a single snapshot observation, centered at a right ascension (RA) of 75.22$^\circ$ and a declination (Dec) of $-30.70^\circ$, inspired by parameters typical of observations by the HERA telescope, \citep{HERA:deboer2017hydrogen,HERA_PhaseII_Berkhout_2024}. 

The map's angular extent is set by the size of the primary beam's main lobe at the longest wavelength, rounded up:
\begin{equation}\label{eq:airy}
    \theta_\mathrm{Airy} \approx 1.22\frac{\lambda_\mathrm{max}}{D} \approx 17^\circ,
\end{equation}
where $\theta_\mathrm{Airy}$ is the angular size of the Airy beam, defined by the full width at its first null, and $\lambda_\mathrm{max} \approx 2.31\,\mathrm{m}$ is the wavelength at 130\,MHz. Conversely, the map resolution is determined by the smallest resolvable angular scale, computed from \autoref{eq:airy} using the shortest wavelength (2\,m) and the longest baseline of the $uv$-complete region (200\,m), yielding a pixel size of 0.283$^\circ$. %During the 10.7-second integration, the sky drifts by approximately 0.045$^\circ$, or about 15\% of the pixel size. This level of drift is considered negligible for the purposes of this demonstration.
Simulation parameters are summarized in \autoref{tab:sim_params}. No noise was added to the visibilities.

\begin{table}[b]
    \centering
    \begin{tabular}{cc}
      %\hline
       %Parameter  & Value \\
       \hline
       Sky models & GLEAM, GSM08\\
       Band  & 130--150\,MHz\\
       Frequency step & 0.5\,MHz\\
       Beam & 10\,m Airy disk\\
       %Integration time & 10.7\,s\\
       RA center & 75.22$^\circ$\\
       Dec center & -30.70$^\circ$\\
       Map RA/Dec range & 17$^\circ$\\
       Map pixel size & 0.283$^\circ$\\
       \hline
    \end{tabular}
    \caption{Simulation and mapping parameters. Only one time observation time was simulated.}
    \label{tab:sim_params}
\end{table}

Finally, the DOM formalism \citep{Xu2022Oct} converts the visibility data into three-dimensional maps using a maximum likelihood estimator for the sky brightness at arbitrary pixel locations, allowing them to be aligned along the frequency axis and enabling a power spectrum estimation with a simple three-dimensional fast Fourier transform.

\begin{figure*}[t]
    \centering
    \includegraphics[width=\textwidth]{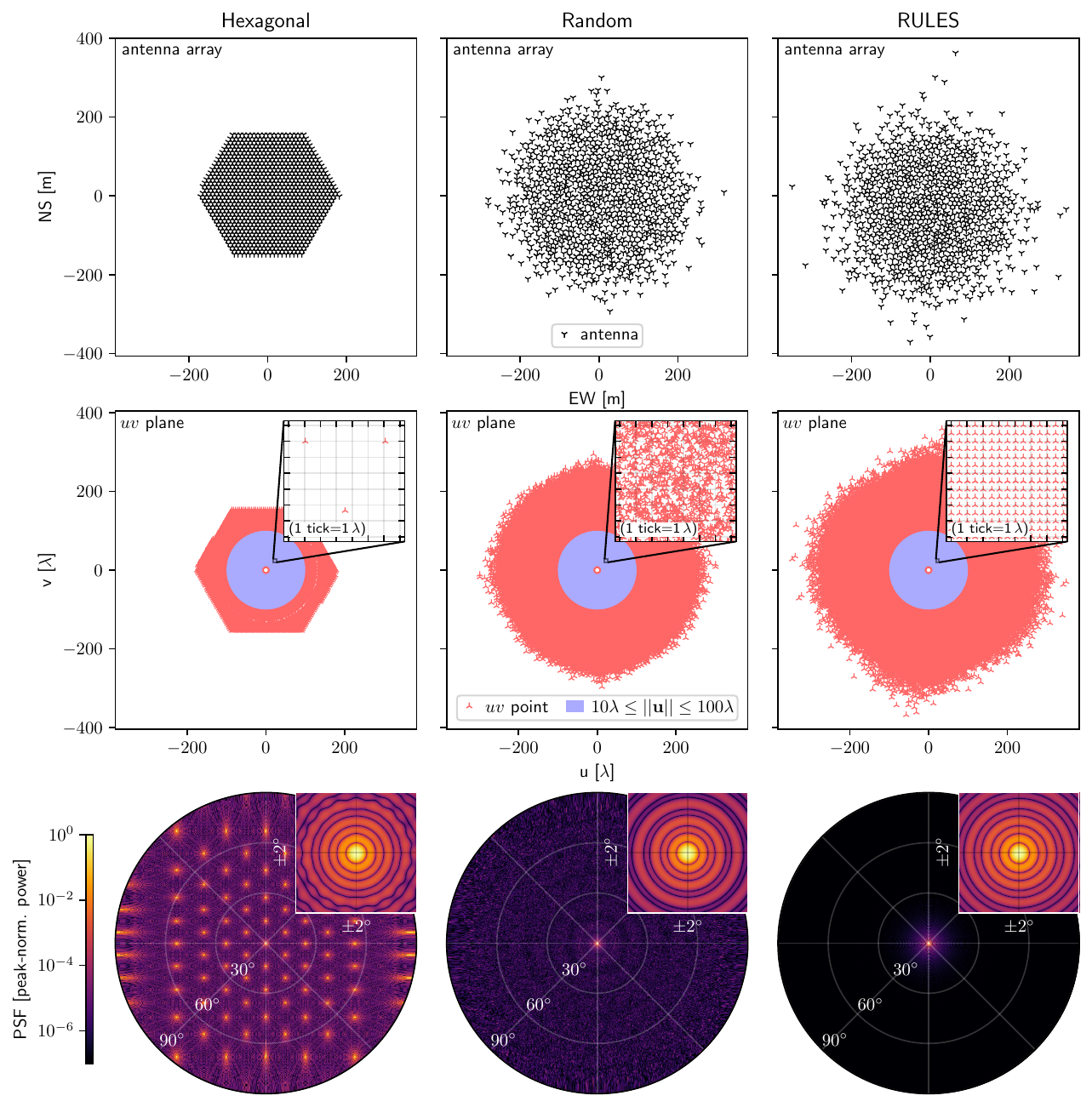}
    \caption{The three arrays used for the simulations in \autoref{sec:wedge}: a regular, close-packed hexagonal array (\texttt{hexagonal}) in the left column, a random array (\texttt{random}) in the central column, and an array generated with the RULES algorithm (\texttt{RULES}) in the right column. The antenna positions are in the first row, $uv$ samples in the second row---with the range of completeness in blue---and the peak-normalized synthesized beams, using only the $uv$ complete range, uniformly weighted, and ignoring primary beam attenuation, in the third row. %The synthesized beam for \texttt{hexagonal} shows dozens of aliased beams within the sky, as expected. The synthesized beam for \texttt{random} shows no alias due to the very dense $uv$ plane, but there is more power all the way out to the horizon. The synthesized beam for \texttt{RULES} shows the first diffraction null at the horizon, and no alias within the sky, as desired. Azimuthally averaged plots for \texttt{random} and \texttt{RULES} are included in \autoref{fig:imaging}.%
    }
    \label{fig:array_uv_beam}
\end{figure*}

Arrays used in the simulations are shown in \autoref{fig:array_uv_beam}, along with their $uv$ coverages and peak-normalized synthesized beams (i.e., PSFs), ignoring primary beam attenuation. Only baselines in the range $10\lambda \leq \norm{\mathbf{u}} \leq 100\lambda$---the region of completeness for \texttt{RULES}---were included when computing all PSFs and power spectra. Including shorter or longer baselines would introduce small-scale features in the PSFs that are unrelated to \texttt{RULES}’s behavior in the target spatial regime, and would extend the power spectrum to modes where we do not claim completeness and thus wedge suppression. Only one baseline per redundant group was simulated, and each was given equal weight, irrespective of redundancy; this uniform weighting scheme, although suboptimal for noise reduction, is necessary for wedge removal. The first array, \texttt{hexagonal}, is a close-packed hexagonal grid designed to represent an extreme case of redundancy-focused layout with minimal $uv$ coverage. While its geometry is inspired by the HERA telescope, it omits HERA’s offset sub-arrays and longer baselines, resulting in even sparser $uv$ sampling. Its synthesized beam exhibits bright grating lobes. The second array, \texttt{random}, provides dense but irregular $uv$ coverage. Its synthesized beam does not have the characteristic grating lobes, but still shows significant power extending to the horizon. The third array, \texttt{RULES}, is the one shown in \autoref{fig:array_layouts}, generated using the RULES algorithm. It achieves a clean synthesized beam with its first diffraction null at the horizon and coherently suppressed power within the sky.

\subsection{Wedge suppression}\label{subsec:ps_results}

\begin{figure*}[t]
    \centering
    \includegraphics[width=\textwidth]{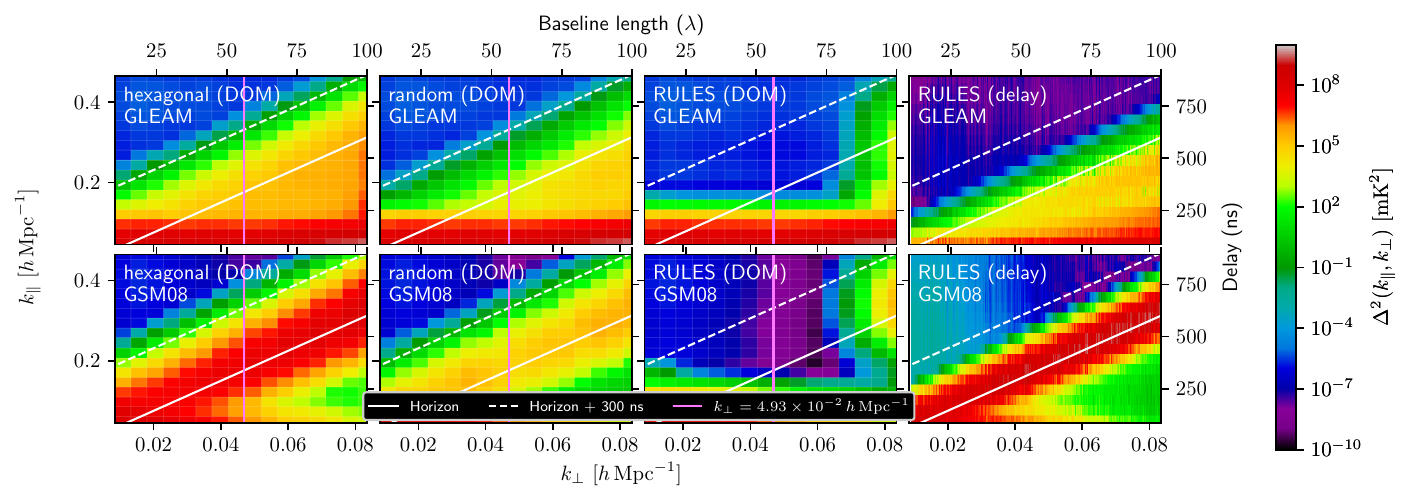}
    \caption{Power spectra for the three arrays presented in \autoref{fig:array_uv_beam}, using visibilities simulated with \texttt{pyuvsim} and the parameters shown in \autoref{tab:sim_params}, then sent through the DOM-PS (first three columns) and delay spectrum (fourth column) pipelines---for a consistency check, a similar analysis with the FHD/$\varepsilon$ppsilon estimator \citep{FHD2012} is shown in \autoref{apsec:eppsilon}. The rows show alternate skies: point sources from GLEAM (top), and diffuse emission from GSM08 (bottom). The horizon and a 200 ns buffer are indicated in a solid and dashed line, respectively. A pink line identifies the $4.93\times10^{-2}h/\mathrm{Mpc}$ cuts shown in \autoref{fig:2dcuts_poserror} and \autoref{fig:2dcuts_rho}. The hexagonal array produces a foreground wedge. The random array shows limited wedge suppression ($\lesssim 3$ orders of magnitude), while the RULES-based array has significant wedge suppression ($\approx 16$ orders of magnitude). Foreground power appears at baseline lengths between 75 and $100\lambda$ because $uv$ completeness is defined at the highest frequency (shortest wavelength) of the band; at 130\,MHz, the array is not complete up to $100\lambda$. The delay power spectrum shows a bright wedge, even with the $uv$-complete array.}
    \label{fig:pspecs}
\end{figure*}

In \autoref{fig:pspecs}, the power spectra for each array are presented for the two different skies (GLEAM for point sources, GSM08 for the diffuse sky). The first three columns are computed with the DOM-PS pipeline \citep{Xu2022Oct,Xu2024Aug}, which uses an $\eta$-transform and can thus suppress wedge power. As expected, the hexagonal array presents a bright foreground wedge, which is only suppressed by up to three orders of magnitude (from $10^8$ to $10^5$\,mK$^2$) in the random realization. Meanwhile, the RULES-based $uv$-complete array exhibits wedge suppression by nearly sixteen orders of magnitude (from $10^8$ to $10^{-8}$\,mK$^2$), essentially hitting the dynamic range of the analysis pipeline. The last column also uses the \texttt{RULES} array, but computes the power spectrum using HERA's delay spectrum estimator, which performs a $\tau$-transform and cannot remove the wedge at any $uv$ density unless the foregrounds are preliminarily subtracted \citep{HERA:deboer2017hydrogen,HERA_PhaseII_Berkhout_2024}. It is included for comparison purposes.

In \autoref{apsec:eppsilon}, we repeat this analysis with the FHD/$\varepsilon$ppsilon estimator \citep{Barry2019Jan} for a consistency check, and find similar result, although we note differences in the power spectra between DOM-PS and FHD/$\varepsilon$ppsilon.

\subsection{Detection of the 21\,cm signal with a realistic array}

While \autoref{subsec:ps_results} shows that RULES-based arrays can suppress the wedge by sixteen orders of magnitude, we have assumed ideal conditions and ignored practical engineering constraints. Previous studies have shown that even very small real-world imperfections can cause significant foreground contamination in the EoR window \citep{orosz2019,Kim2023Aug}. Since RULES achieves wedge suppression through careful baseline selection, we investigate what happens when those baselines are either perturbed (from antenna position errors) or missing (e.g., due to hardware failures). To assess whether these effects compromise the detection of the 21\,cm signal, we compare the resulting power spectra to those from a pure HI simulation generated using \texttt{21cmFAST} \citep{21cmfast2011,21cmfast2020} with the fiducial EoR model from \cite{Park2019_fiducial_EoR}; the resulting simulated coeval cubes were tiled to form a full sky model at the redshifts of interest using the technique described in the appendices of \cite{kittiwisit2017} and then sent through the same pipeline as the foreground models, assuming the unperturbed \texttt{RULES} array.

\begin{figure}
    \centering
    \includegraphics[width=\linewidth]{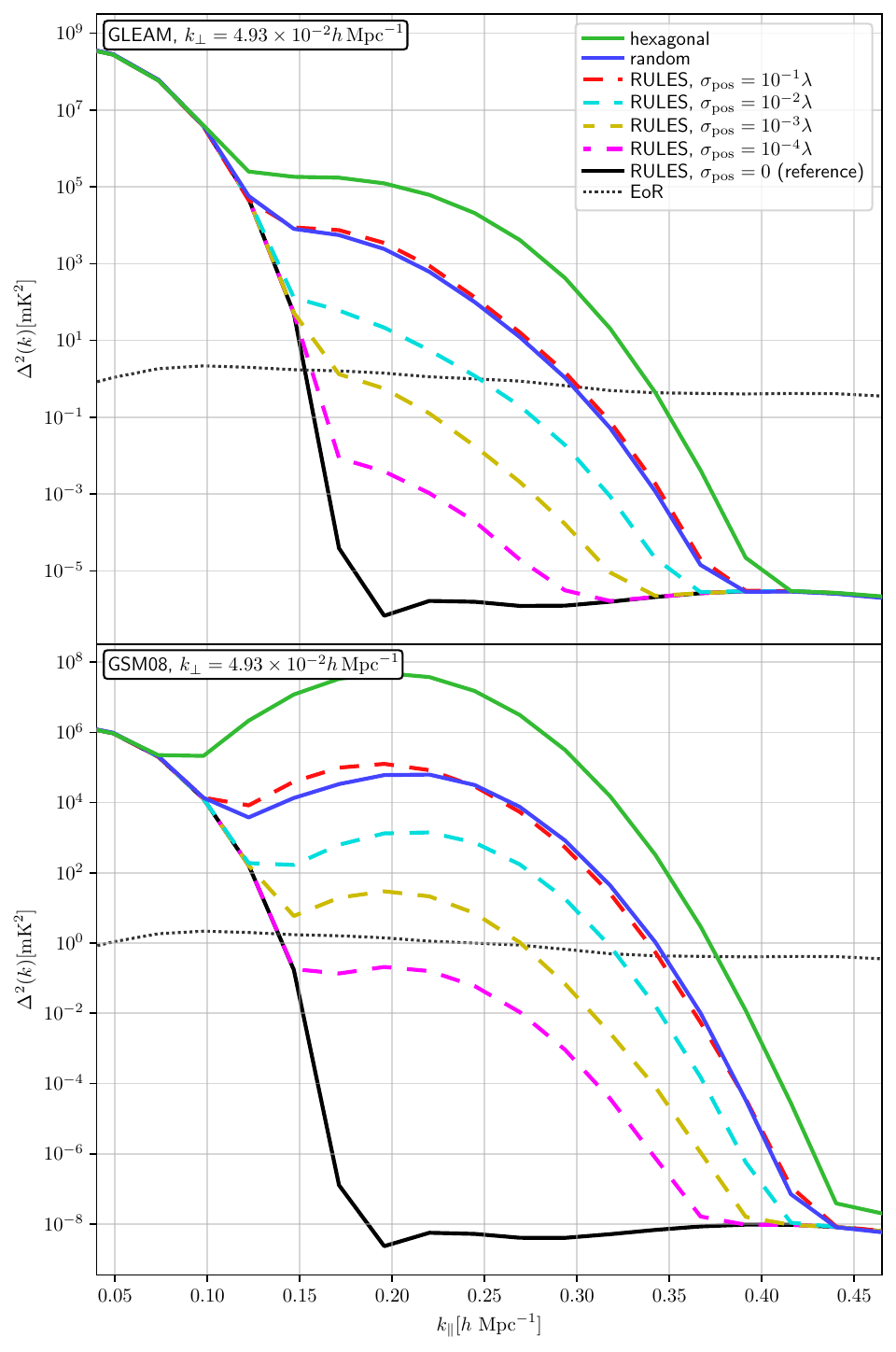}
    \caption{Cuts through the 2D power spectrum for arrays where random antenna displacements along the EW and NS directions, up to $\sigma_\mathrm{pos}$, were applied. The reference wavelength $\lambda$ is 2\,m. Even with a very lenient tolerance of 20\,cm, the RULES realization outperform random arrays; a stricter but still realistic tolerance of 2\,mm suppresses the wedge by more than 5 orders or magnitude.}
    \label{fig:2dcuts_poserror}
\end{figure}

We first examine the impact of antenna position errors by defining a maximum error $\sigma_\mathrm{pos}$ and applying random displacements to each antenna in \texttt{RULES}, uniformly sampled from $[-\sigma_\mathrm{pos},\sigma_\mathrm{pos}]$ in both the EW and NS directions (no displacement along the up-down axis). The perturbed arrays are processed through the same simulation and power spectrum pipeline as in \autoref{subsec:ps_results}. \autoref{fig:2dcuts_poserror} shows results for a cut at $k_\perp=4.93\times 10^{-2} h\mathrm{Mpc}^{-1}$ (indicated by the pink line in \autoref{fig:pspecs}), chosen as a representative bin within the wedge region, along with the fiducial EoR model. This reveals a strict tolerance requirement: even with $\sigma_\mathrm{pos} = 10^{-2}\lambda = 2\,\mathrm{cm}$, wedge power significantly exceeds the 21\,cm line at most $k_\parallel$, though it still outperforms a random array by approximately two orders of magnitude. Improving precision beyond this point yields substantially more wedge suppression, nearly reaching EoR levels everywhere at $\sigma_\mathrm{pos}\lesssim 10^{-3}\lambda = 2\,\mathrm{mm}$. The discretized aperture plane geometry (see \autoref{subsec:redundancy}) may facilitate achieving such precise positioning, but we recognize that this sensitivity to small displacements is a significant practical challenge, particularly for high-frequency arrays (e.g. post-reionization, near 1\,GHz); experiments observing at longer wavelengths will face a less stringent requirement.

\begin{figure}
    \centering
    \includegraphics[width=\linewidth]{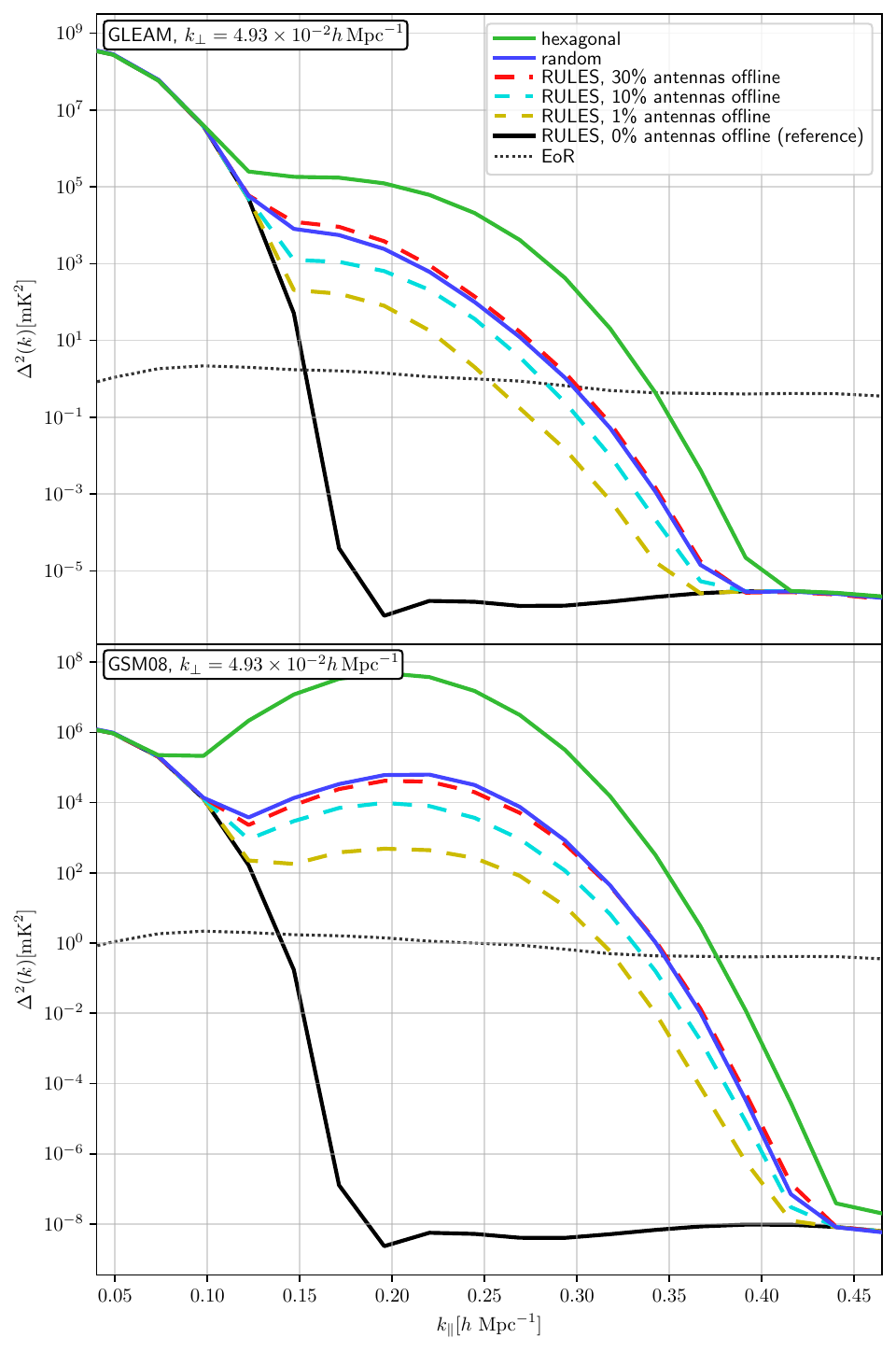}
    \caption{Cuts through the 2D power spectrum for arrays where a random subsample of antennas are removed. Wedge suppression is very sensitive to missing antennas, considering that more than ten orders of magnitude of foreground power come back when only 1\% of antennas are missing. This could be remediated by imposing a minimum redundancy requirement. Only when 30\% of antennas are missing does the RULES-based array still perform like a random realization.}
    \label{fig:2dcuts_mf}
\end{figure}

The second feasibility test examines performance when antennas go offline, such as during hardware failures or maintenance operations---a routine occurrence in any observatory. We simulated the RULES array with randomly selected antennas removed and processed the degraded arrays through our standard pipeline. While the built-in exact redundancy (see \autoref{subsec:redundancy}) means that removing a single antenna does not necessarily eliminate all associated $uv$ samples---since other baselines may provide a redundant measurement---the results in \autoref{fig:2dcuts_mf} reveal extreme sensitivity to this failure mode. Even with only 1\% of antennas offline, foreground power in the wedge raises above the 21\,cm signal. This vulnerability can be preempted by designing RULES arrays with minimum redundancy requirements for each baseline, ensuring $uv$-coverage completeness despite antenna failures. When we impose a minimum twofold redundancy constraint on an array with identical parameters to \autoref{fig:array_layouts}, the required antenna count increases to 1,285; for fivefold minimum redundancy, this rises to 2,044 antennas. These numbers remain practically feasible and scale slowly with the redundancy requirement, suggesting a viable path toward robust RULES implementations.

\begin{figure}
    \centering
    \includegraphics[width=\linewidth]{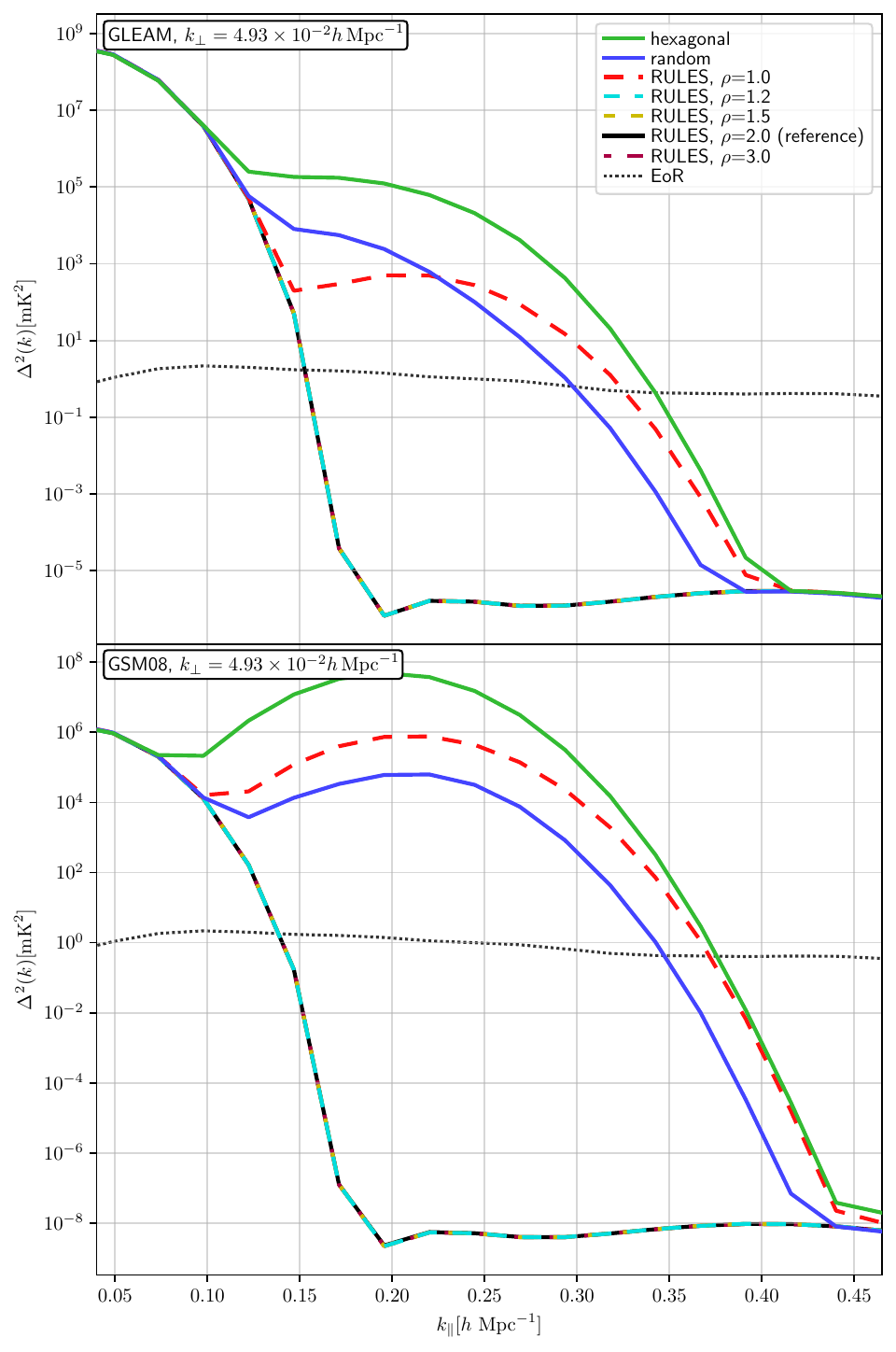}
    \caption{Cuts through the 2D power spectrum for RULES-based arrays with regular $uv$ coverage different values of $\rho$, where $\rho$ is the sampling density in the $uv$ plane (one point per $\lambda/\rho$-sized square cell). In both cases, all densities but $\rho=1$ are indistinguishable from the reference.}
    \label{fig:2dcuts_rho}
\end{figure}

An additional consideration for feasibility is whether our completeness criterion could be relaxed to allow regular but less dense $uv$ coverage, thereby requiring fewer antennas. We generated and simulated arrays with $\rho=1$, 1.2, and 1.5 with results shown in \autoref{fig:2dcuts_rho}. We find that all arrays with $\rho \geq 1.2$ produce indistinguishable power spectra, while the $\rho = 1$ case performs similarly---if slightly worse---to random arrays.

\begin{figure}
    \centering
    \includegraphics[width=\linewidth]{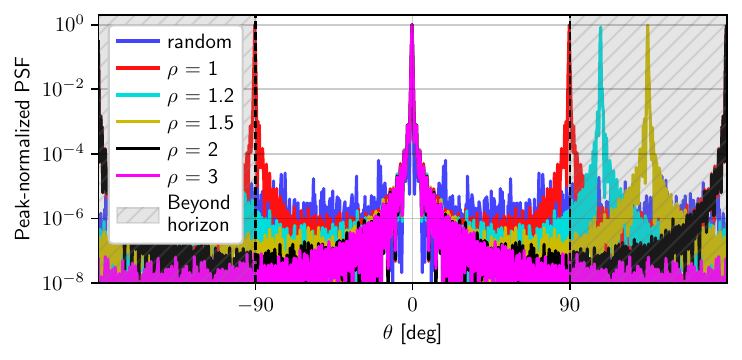}
    \caption{Cuts through the peak-normalized PSF for the \texttt{random} array, along with three RULES-based array at different values of $\rho$. In the relatively sparse cases of $\rho=1.5$ and even $\rho=1.2$, the first diffraction peak in the PSF is mostly beyond the horizon, and the in-sky power is coherently suppressed at the horizon by a few orders of magnitude more than \texttt{random}; in comparison, $\rho=1$ sees its power peak again precisely at the horizon.}
    \label{fig:rho_leq_2}
\end{figure}

This suggests that our original completeness criterion---which assumed uniform sky sensitivity out to the horizon---may have been overly strict. In reality, the primary beam attenuates emission near the horizon, effectively reducing the angular extent of the observable sky and thus relaxing the required $uv$ sampling density; the coherent suppression from regular sampling provides most of the wedge suppression. Put differently, as $uv$ coverage becomes sparser, the PSF's diffraction pattern narrows and the first diffraction peak moves inward. For moderate reductions in $\rho$, the power that re-enters the visible sky remains sufficiently faint to be strongly suppressed by the primary beam. This heuristic is illustrated in \autoref{fig:rho_leq_2}.

\begin{figure}
    \centering
    \includegraphics[width=\linewidth]{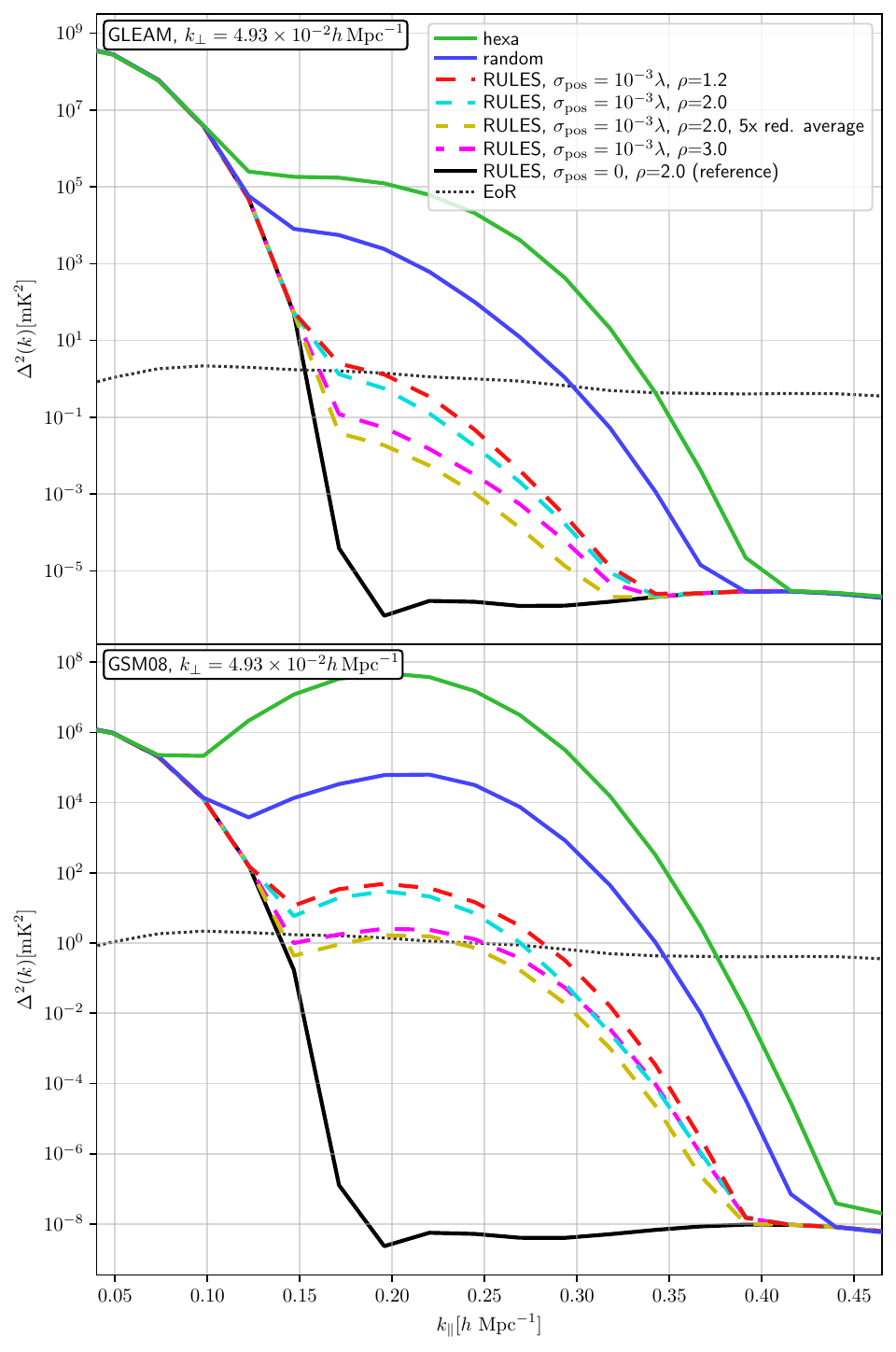}
    \caption{Cuts through the 2D power spectrum for arrays with position error $\sigma_\mathrm{pos}=10^{-3}\lambda=2\text{\,mm}$ for different $uv$ packing densities $\rho$. While in the absence of position errors, the power spectra for $\rho=1.2$, 2, and 3 are identical (see \autoref{fig:2dcuts_rho}), introducing such errors show how a tighter $uv$ packing is favorable. An additional cut shows the power spectrum for the $\rho=2$ case, but using an array with a fivefold minimum redundancy requirement, and redundantly averaging the visibilities before sending them through the DOM-PS pipeline; this approach also helps mitigate the effects of position errors.}
    \label{fig:2dcuts_realistic_Delta_squared}
\end{figure}

This is an important finding because it demonstrates that arrays with moderately lower $uv$ densities---requiring fewer antennas---can achieve comparable performance. We do however find that sparser arrays perform slightly worse when combined with antenna position errors; conversely, $uv$ densities beyond $\rho\geq 2$ appear to help. This is illustrated in \autoref{fig:2dcuts_realistic_Delta_squared}, where we also added the curve for an array with $\rho=2$ and a fivefold minimum redundancy requirement, where the visibilities were redundantly averaged before sending them through the DOM-PS pipeline, which shows attenuated foreground power inside the wedge. This indicates that leakage caused by position errors could be mitigated by a higher value of $\rho$ or a higher redundancy count, both requiring more antennas; meanwhile, an array with very low position error could use much fewer antennas by reducing $\rho$.

\section{Discussion}\label{sec:discussion}

\subsection{Importance of $uv$ completeness}

Working within the region obscured by the wedge is essential for imaging-based 21\,cm science, such as directly reconstructing the 21\,cm field or cross-correlating with galaxy surveys, since excluding an asymmetric portion of Fourier space fundamentally prevents image reconstruction \citep{beardsley_2015, cross_cor_wedge_2016, cross_corr_cohn_2016, cross_cox_2022, gagnon_hartman_2024}. It has also been shown that existing wedge removal processes tend to destroy the information content of one-point statistics \citep{kittiwisit2017, Kim2025_onepoint}. Additionally, those newly-unlocked wedge regions---at lower $k$ modes---correspond to large spatial scales where the 21\,cm signal is intrinsically stronger. To date, no analysis method in the field has succeeded in recovering wedge modes at the dynamic range required for 21\,cm cosmology, motivating layout-based approaches such as this one or that of \cite{MurrayTrott2018Dec}.

Arrays with complete $uv$ coverage also help with calibration, where even the smallest errors can flood the cosmological window in bright foregrounds \citep{calib_barry_2016}. By sampling a larger number of independent modes, $uv$-complete arrays allow for a more exhaustive comparison between the measured sky and the calibration sky model, enabling more accurate calibration than is possible with redundant calibration on regular arrays \citep{calib_Byrne_2019}. Furthermore, the suppression of the wedge reduces spectral calibration errors induced by unmodeled foregrounds, such as faint sources absent from the calibration catalog \citep{calib_ewallwice_2017}.

As noted in \autoref{subsec:redundancy}, $uv$-complete arrays, while more redundant than random configurations, remain markedly less redundant than highly regular layouts such as HERA \citep{HERA:deboer2017hydrogen} or CHORD \citep{Vanderlinde2019Oct}. Lower redundancy implies higher thermal noise in power spectrum estimates due to reduced coherent averaging, a problem that is aggravated by the fact that wedge suppression from complete coverage only works with uniform baseline weighting. However, this trade-off can be mitigated by longer integration times or by imposing a minimum redundancy requirement, and may be offset by the increased number of usable modes and improved calibratability. Quantifying these trade-offs---between thermal noise, cosmological sensitivity, and calibration performance under realistic assumptions (number and size of antennas, instrumental noise, bandwidth, etc.), in the spirit of \cite{Pober2014ApJ}, is left to future work. Here, we focus on demonstrating the feasibility of $uv$-complete arrays, presenting the generating algorithm, and highlighting how such arrays can eliminate the foreground wedge.

\subsection{High resolution imaging beyond 21 cm}

The PSFs in \autoref{fig:array_uv_beam} show that \texttt{RULES} achieves significantly stronger suppression outside of the main lobe compared to \texttt{random}---by several orders of magnitude---raising the question of whether such arrays might also be advantageous for science goals beyond 21\,cm cosmology, that also necessitate well-behaved PSFs but additionally require high angular resolutions, such as those pursued by the DSA-2000 \citep{DSAHallinan2019Sep}. However, a fundamental limitation remains: the longest baselines in \texttt{RULES} are relatively short, and this is a necessary feature of $uv$-complete arrays, as compact layouts increase the chance of fulfilling multiple different commanded baselines simultaneously. This is in direct tension with the specifications of imaging-focused observatories, which, to achieve high angular resolution, require long baselines. For example, the DSA-2000 will span baselines up to 15\,km, corresponding to a resolution of $\sim$\,0.07\,arcmin at 1\,GHz. Achieving $uv$ completeness across that range would require $\sim 1.5 \times 10^{10}$ $uv$ samples. Even under ideal conditions---in which every single baseline is unique and fulfills a distinct commanded point---this would necessitate nearly $2 \times 10^5$ antennas. A collaboration that is planning to build an imaging array with $N_\mathrm{A}\gtrsim\mathcal{O}(10^3)$ may nonetheless want to consider using a fraction of their antennas as a $uv$-complete core to complement their sparser long baselines and unlock possibilities for a secondary 21\,cm science goal.

\subsection{More efficient solutions}

The RULES algorithm demonstrates the feasibility of $uv$-complete arrays under realistic geometric constraints and costs, but does not claim to produce layouts that use the minimal possible number of antennas for a given set of commanded points. Indeed, even if all possible $\{\mathbf{a}_\mathrm{ref}, \mathbf{u}_\mathrm{C}\}$ pairs are evaluated at each iteration---a computationally intensive approach---an even more optimal placement for a given antenna may still be revealed after subsequent antennas are added. A potentially more effective, though significantly costlier, strategy would involve evaluating combinations of \textit{more than one} antenna at each step and selecting the configuration that maximizes the number of newly fulfilled $uv$ points. Now that the viability of $uv$-complete layouts under practical constraints has been demonstrated, future work can focus on discovering more economical generating strategies that achieve the same coverage with fewer antennas.

A promising direction for future work is to draw on the mathematical literature to develop improved algorithms or formal definitions of optimality. Arrays that achieve $uv$ completeness with the fewest possible antennas are conceptually related to combinatorial structures known as Golomb arrays (or their close cousin, Costas arrays), which avoid repeated pairwise separations in $k$ dimensions (whereas RULES tolerates those repetitions and instead focuses on realizing \textit{all} pairwise separations within some range). The better-known one-dimensional variant, the Golomb \textit{ruler}, has been considered in earlier generations of radio telescopes \citep{bireau1974golomb}. These mathematical objects come with well-defined optimality criteria and established construction algorithms. While most recent work has focused on one-dimensional applications \citep{ogr_apr2021,ogr_july2021,ogr2024}, extensions to two dimensions have also been explored \citep{costas1984,Golomb2d2000}, including in the context of $uv$ sampling for radio interferometry \citep{ebrahimi_may2023,lazko_nov2023}. These latter efforts, however, have typically been limited to relatively small numbers of antennas and $uv$ packing densities much lower than $\rho=2$ such that collisions were not a limiting factor. Yet, combining these advances to the $uv$ completeness criterion introduced in this work and physical collision constraints could lead to new algorithmic strategies, or even formal proofs of the minimal antenna count required under realistic design considerations.

\section{Conclusion}\label{sec:conclusion}

We have defined a $uv$ completeness criterion and presented the RULES algorithm for constructing antenna arrays that satisfy this definition within a specified range of baseline lengths, under realistic constraints. RULES incrementally builds the array by placing antennas to fulfill a target set of $uv$ points, selecting each placement to maximize the number of newly fulfilled points. We showed that complete $uv$ coverage over the $10\lambda\leq\norm{\mathbf{u}}\leq 100\lambda$ range with antennas of diameter $5\lambda$ is achievable with less than 1000 antennas, consistent with current designs for prospective instruments.

The primary motivation for this work is the suppression of the foreground wedge in 21\,cm power spectrum analyses. We performed noiseless visibility simulations over the 130--150\,MHz band using foreground-like sky models and three array types: a regular but $uv$-sparse layout, a random $uv$-dense layout, and an RULES-based $uv$-complete layout. We then computed the corresponding power spectra with image-based estimators. The $uv$-sparse array exhibits a bright wedge; the random array shows up to three orders of magnitude of wedge suppression compared to the $uv$-sparse case; the $uv$-complete array achieves sixteen orders of magnitude of suppression. This result is sensitive to small antenna position errors, presenting an engineering challenge and suggesting that $uv$ complete layouts may be preferred for longer-wavelength applications. We propose ways to mitigate this issue, namely increasing the $uv$ packing density or the redundancy count---which both require more antennas---while leveraging the fact that antennas are located on a discretized grid to help achieve a strict position tolerance. We also showed that the results hold for sparser---but still regular---$uv$ coverages, but are very sensitive to missing antennas, a problem that could also be addressed by increasing the redundancy count. Even in the worst-case scenarios, $uv$ complete arrays perform at least as well as random arrays with the same number of antennas and physical footprint, and improvements on that worst-case scenario suppress the wedge by many orders of magnitude, potentially well below EoR levels. These results demonstrate that $uv$-complete arrays are theoretically well-motivated and provide substantial benefits even in non-ideal implementations.

\section*{Acknowledgements}

We thank Bryna Hazelton and Honggeun Kim for their assistance with the FHD/$\varepsilon$ppsilon and \texttt{21cmFAST} softwares, respectively, and Tyler Cox, Joshua Dillon, Miguel Morales, and Steven Murray for insightful discussions that helped shape this paper. V.M. and J.N.H. gratefully acknowledge support from the MIT School of Science and the Gordon and Betty Moore Foundation (the latter through grant GBMF5212 to the Massachusetts Institute of Technology). R.B. is supported by the National Science Fundation Award No. 2303952.

\section*{Software Availability}\label{sec:software_availability}

All softwares used in this paper are available publicly, starting with the RULES algorithm itself.\footnote{\url{https://github.com/vincentmackay/uvrules}} The visibility simulations were computed with \texttt{pyuvsim}\footnote{\url{https://github.com/RadioAstronomySoftwareGroup/pyuvsim}} \citep{pyuvsim_Lanman2019}, while the power spectra were computed using the DOM\footnote{\url{https://github.com/HERA-Team/direct_optimal_mapping}} \citep{Xu2022Oct,Xu2024Aug}, FHD/$\varepsilon$ppsilon\footnote{\url{https://github.com/EoRImaging/FHD}\\\url{https://github.com/EoRImaging/eppsilon}} \citep{Barry2019Jan}, and \texttt{hera\_pspec}\footnote{\url{https://github.com/HERA-Team/hera_pspec}} \citep{HERA:deboer2017hydrogen,HERA_PhaseII_Berkhout_2024} frameworks. The \texttt{21cmFast}\footnote{\url{https://github.com/21cmfast/21cmFAST}} \citep{21cmfast2011,21cmfast2020} simulations were tiled with the \texttt{cosmotile}\footnote{\url{https://github.com/steven-murray/cosmotile}} package \citep{kittiwisit2017}.

\appendix

\section{Other algorithmic parameters}\label{apsec:parsweeps}

\begin{figure*}
    \gridline{
        \fig{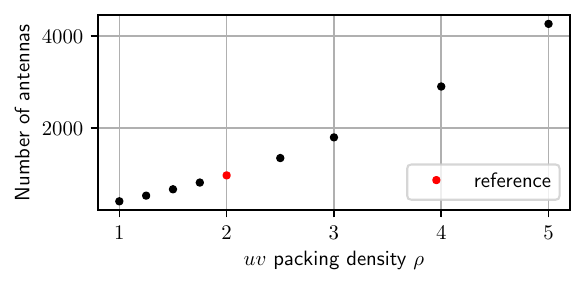}{0.48\textwidth}{(a)}
        \fig{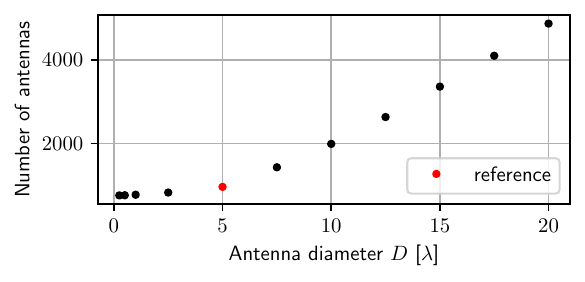}{0.48\textwidth}{(b)}
    }
    \vspace{-0.5em}
    \gridline{
        \fig{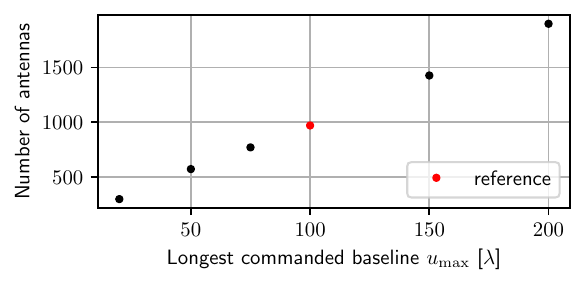}{0.48\textwidth}{(c)}
        \fig{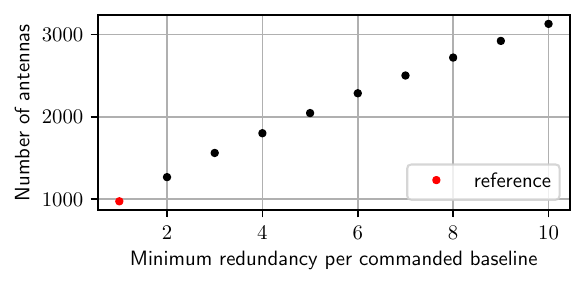}{0.48\textwidth}{(d)}
    }
\caption{Number of antennas required to fulfill all commanded baselines as a function of (a) the $uv$ packing density $\rho$, (b) the dish diameter used as a collision constraint, (c) the maximum allowed baseline length, and (d) the minimum redundancy number of each commanded baseline.}
\label{fig:parsweeps}
\end{figure*}

The RULES-based array used throughout this paper was generated using the commanded baselines described in \autoref{subsec:commanded_baselines}, with a minimum antenna spacing set by the antenna size of $D=5\lambda$ and $uv$ packing density $\rho = 2$. The maximum commanded baseline length $u_\mathrm{max}$ is $100\lambda$, and the minimum redundancy requirement for each commanded baseline was one. To assess the algorithm’s sensitivity to these parameters, we vary each one independently and present the results in \autoref{fig:parsweeps}. We recognize the degeneracy between some of those parameters: in units of wavelength, an array layout will be identical if $\rho$ is halved, but $D$, $u_\mathrm{min}$, and $u_\mathrm{max}$ are doubled. Nonetheless, we present those parameters independently as it is more intuitive.

We find that within the ranges tested, the number of antennas scales almost linearly with both $\rho$ and $u_\mathrm{max}$, as shown in \autoref{fig:parsweeps}a and \autoref{fig:parsweeps}c. This trend is expected: the number of commanded $uv$ points grows like the square of these parameters, while the number of baselines also scales quadratically with the number of antennas. If all new baselines created by introducing a new antenna only fulfills yet unfulfilled commanded points, this would result in a perfectly linear relationship. A more peculiar behavior is seen in \autoref{fig:parsweeps}b where there appears to be two linear regimes; a moderate slope for $D\leq 5\lambda$, that gets much steeper for $D\geq 5\lambda$. This is likely just a feature of the range covered. Indeed, going from $D=5\lambda$ to $D=20\lambda$ is similar to increasing $\rho$ by a factor 4. While a given observatory would likely see no interest in going much above $\rho=2$, it is not uncommon that modern arrays have $D\geq 20\lambda$. \autoref{fig:parsweeps}d shows how the minimum redundancy per commanded baseline affects the number of antennas; the relationship seems to be linear, costing approximately 220 new antennas per redundancy number. This is a relatively slow growth considering that the reference case already requires 971 antennas: by doubling the number of antennas, we obtain a five-fold increase in redundancy, which provides benefits such as increased sensitivity and robustness to antennas going offline.

\begin{figure*}
    \gridline{
      \fig{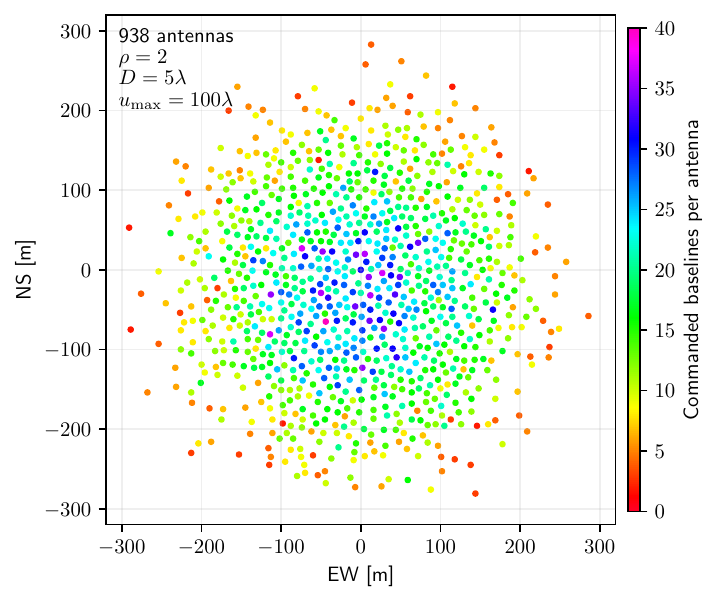}{0.48\textwidth}{(a)}
      \fig{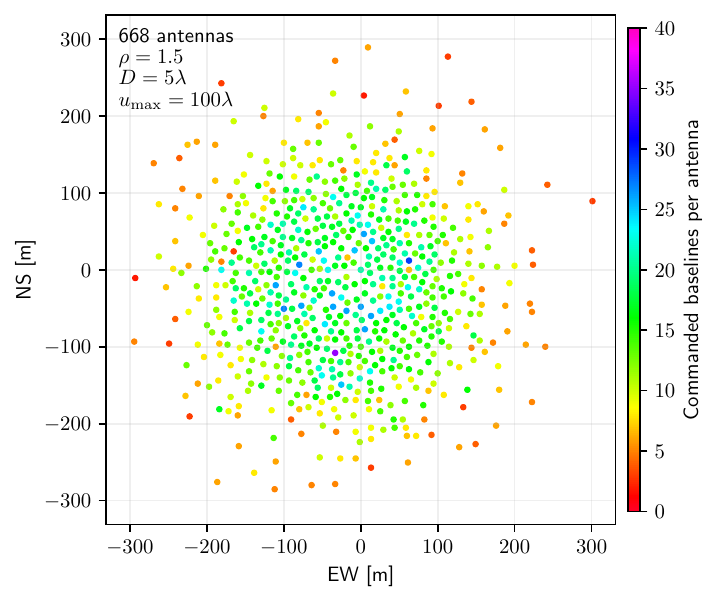}{0.48\textwidth}{(b)}
    }
    \vspace{-0.5em}
    \gridline{
      \fig{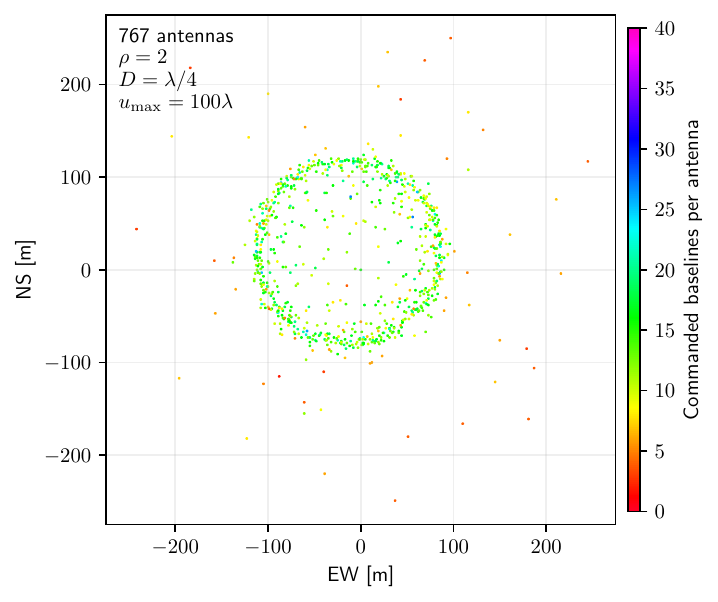}{0.48\textwidth}{(c)}
      \fig{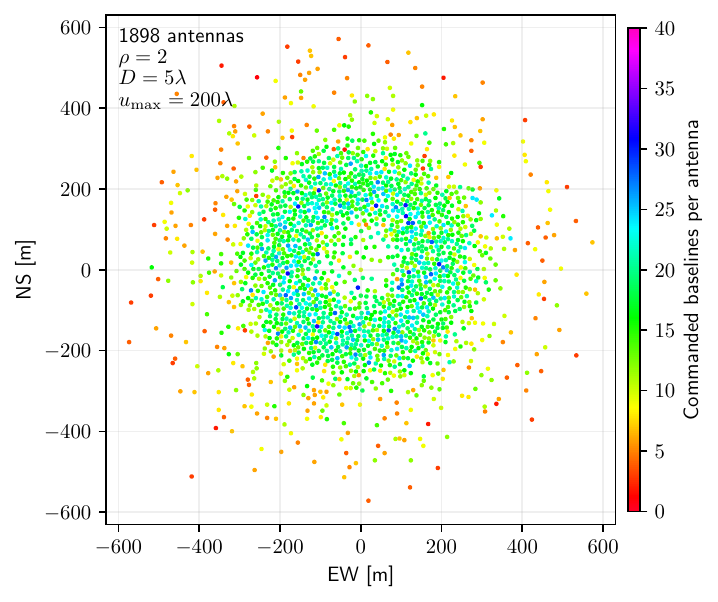}{0.48\textwidth}{(d)}
    }
    \caption{Additional array layouts. (a) is the array generated with the same parameters as in \autoref{subsec:the_algorithm}, but comparing all $\{\mathbf{a}_\mathrm{ref},\mathbf{u}_\mathrm{C}\}$ pairs at each iteration, which is computationally costly but requires fewer antennas. The three other arrays vary only one parameter; (b) has $uv$ packing density $\rho=1.5$; (c) uses a dish diameter of $D=\lambda/4$ as the collision constraint; (d) has a maximum commanded baseline length $u_\mathrm{max}=200\lambda$.}
    \label{fig:extra_arrays}
\end{figure*}

In \autoref{fig:extra_arrays}, we present some select array layouts. \autoref{fig:extra_arrays}a is the array generated with the same parameters as in \autoref{subsec:the_algorithm}, but comparing all $\{\mathbf{a}_\mathrm{ref},\mathbf{u}_\mathrm{C}\}$ pair at each iteration, which is very computationally costly, but requires fewer antennas (938 instead of 971). The three other arrays have the same parameters except for one; they represent points from the subplots in \autoref{fig:parsweeps}. \autoref{fig:extra_arrays}b is an array with $\rho=1.5$; \autoref{fig:extra_arrays}c has a dish diameter of $\lambda/4$; \autoref{fig:extra_arrays}d has a maximum commanded baseline length of $200\lambda$.

We note that the arrays shown in \autoref{fig:extra_arrays}c and \autoref{fig:extra_arrays}d exhibit ring-like antenna distributions---a common outcome of RULES, particularly when the ratio $u_\mathrm{max}/D\gtrsim\mathcal{O}(100)$. While we do not offer a definitive explanation for this behavior, it could arise, for example, if the same reference antenna $\mathbf{a}_\mathrm{ref}$ is selected repeatedly over many iterations. A potential direction for future work is to investigate whether distinct algorithmic parameter choices could similarly be associated with characteristic array geometries, and whether such patterns can inform faster construction of $uv$-complete arrays without relying on the full algorithm.

%\section{Generating the random array}\label{apsec:random}

%In \autoref{subsec:redundancy} and \autoref{sec:wedge}, an array generated with the algorithm (\texttt{RULES}) is presented alongside a comparable random array (\texttt{random}). The latter was created via the following process:
%
%\begin{enumerate}
%    \item Place an antenna at (0,0).
%    \item Pick a random position EW and NS position in $[-R,R]$, where $R$ is the maximum radius of the array, using to a normal distribution centered at 0 with some standard deviation $\sigma$. Keep the position if it falls within a radius $R$ of $(0,0)$; pick a new one otherwise.
%    \item Tentatively place an antenna at that position; if there is no collision, add this antenna to the array; otherwise, repeat the previous step.
%    \item Stop when the number of antennas has reached the desired number $N_\mathrm{A}$.
%\end{enumerate}
%
%The values of $R$ and $N_\mathrm{A}$ are trivially picked to be identical to the corresponding values from \texttt{RULES}. However, $\sigma$ cannot simply be cribbed from \texttt{RULES}'s spatial distribution; the algorithm, especially the collision constraint, makes it so that the radii of either arrays are not normally distributed. For simplicity, we thus estimated a value $\sigma$ by eye such that the arrays looked most similar. The resulting \texttt{random} is presented in the top central panel of \autoref{fig:array_uv_beam}, with \texttt{RULES} to its immediate right.

\section{Power spectra using FHD/$\varepsilon$ppsilon}\label{apsec:eppsilon}

\begin{figure}[ht]
    \centering
    \includegraphics[width=\textwidth]{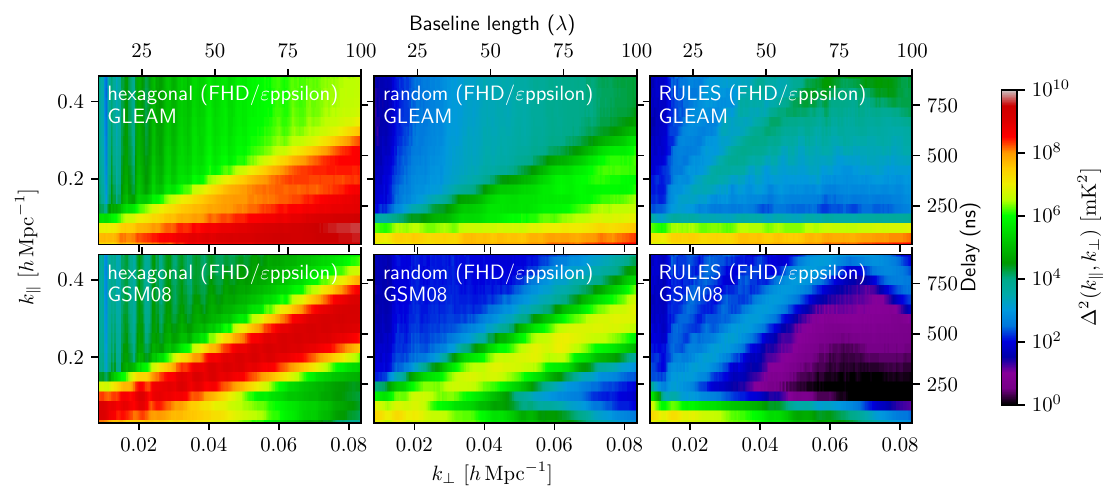}
    \caption{Power spectra for the three arrays presented in \autoref{fig:array_uv_beam}, using visibilities simulated with \texttt{pyuvsim} and the parameters shown in \autoref{tab:sim_params}, and then sent through the FHD/$\varepsilon$ppsilon power spectrum estimator pipeline \citep{Barry2019Jan}, to be compared with \autoref{fig:pspecs}. Here again, the RULES-based array outperforms the random realization in terms of wedge suppression by multiple orders of magnitude. We also note other qualitative differences between these spectra and those shown in \autoref{fig:pspecs}; we leave the investigation of those features to future work.}
    \label{fig:eppsilon}
\end{figure}

As a consistency check, we processed the simulated visibilities through the FHD/$\varepsilon$ppsilon power spectrum pipeline \citep{Barry2019Jan}, and present the resulting spectra in \autoref{fig:eppsilon}. Unlike DOM---which computes the maximum-likelihood value at arbitrary pixel locations and enables a 3D FFT to estimate the power spectrum---FHD grids the data in visibility space. $\varepsilon$ppsilon then performs only a one-dimensional Fourier transform along the frequency axis, before binning the results and performing a weighted average to generate one- and two-dimensional power spectra. While the wedge suppression is not as significant with this pipeline, the $uv$-complete array still achieves over five orders of magnitude of suppression---definitely outperforming a random array---and possibly sufficient for a 21\,cm detection. We do note that the differences with \autoref{fig:pspecs} are not trivial, including a residual wedge-like structure appearing in the \texttt{RULES} power spectrum for both foreground models, raising the question of whether all image-based estimators perform equally under the specific conditions of complete $uv$ coverage---a question we leave to future work.

\bibliography{bibliography}{}
\bibliographystyle{aasjournal}



\end{document}